\documentclass[12pt,epsfig]{article}
\usepackage{amssymb,epsfig}

\makeatletter
\@addtoreset{equation}{section}

\newcommand{\cI}{{\cal I}}
\newcommand{\ScrIP}{{\cI^{+}}}

\newcommand{\cM}{{\cal M}}

\newcommand{\cR}{{\cal R}}

\newcommand{\mh}{{M_H}}

\def\be{\begin{equation}}
\def\ee{\end{equation}}
\def\beq{\begin{equation}}
\def\eeq{\end{equation}}
\def\bea{\begin{eqnarray}}
\def\eea{\end{eqnarray}}
\begin{document}

\begin{titlepage}
\hfill
\vbox{
    \halign{#\hfil         \cr
           } 
      }  
\vspace*{20mm}

\begin{center}
{\Large {\bf On the Topology of Black Hole Event Horizons\\
 in Higher Dimensions
}\\} 
\vspace*{15mm}

{\sc Craig Helfgott, Yaron Oz and Yariv Yanay}

\vspace*{1cm} 
{\it Raymond and Beverly Sackler Faculty of Exact Sciences\\
School of Physics and Astronomy\\
Tel-Aviv University , Ramat-Aviv 69978, Israel\\}

\end{center}

\vspace*{8mm}
 
\begin{abstract}

In four dimensions  the topology of the event horizon of an asymptotically
flat stationary black hole is 
uniquely determined to be the two-sphere $S^2$.
We consider the  topology of event horizons in higher dimensions.
First, we reconsider Hawking's
 theorem and show that the integrated Ricci scalar 
curvature with respect to the induced metric
on the event horizon is positive also in higher dimensions.
Using this and Thurston's geometric types
classification of three-manifolds, we find that the 
only possible
geometric types of event horizons in five dimensions are $S^3$
and $S^2 \times S^1$.
In six dimensions we use the requirement that
the horizon is cobordant to a four-sphere (topological censorship), Friedman's
classification of topological four-manifolds and Donaldson's results
on smooth four-manifolds,  and show that 
simply connected event horizons are homeomorphic to $S^4$ or $S^2\times S^2$. 
We find allowed non-simply connected event 
horizons $S^3\times S^1$ and  $S^2\times \Sigma_g$,
and event 
horizons with finite non-abelian first homotopy group, whose universal
cover is  $S^4$.
Finally, we discuss the classification
in dimensions higher than six.

\end{abstract}
\vskip 0.8cm


\end{titlepage}

\setcounter{footnote}{0}

\section{Introduction}

In four dimensions  the topology of the event horizon of an asymptotically
flat stationary black hole is 
uniquely determined to be the two-sphere $S^2$ \cite{HE} \footnote{
We consider the connected part of the event horizons.}.
Hawking's theorem  \cite{HE} requires 
the integrated Ricci scalar 
curvature with respect to the induced metric
on the event horizon to be positive. This condition
applied to two-dimensional manifolds determines uniquely the topology.

Another way to determine the topology of
the event horizon is via
the  so called topological censorship
\cite{Friedman:1993ty}.
Mathematically it requires the horizon to be cobordant to a sphere
via a simply connected oriented cobordism. 
For a two-dimensional horizon it means
that there is a simply connected three-dimensional oriented manifold
whose boundary is the oriented disjoint union of
the horizon and the two-sphere.
Topological censorship implies that the
 topology of
the event horizon is that of the two-sphere $S^2$ 
\cite{Chrusciel:1994tr}.

The classification of the  topology of the event horizons in higher dimensions
is more complicated \cite{Reall:2002bh}.
For instance, for five-dimensional asymptotically
flat stationary black holes, 
in addition to the known $S^3$ topology of event horizons,
stationary black hole solutions with  event horizons of 
$S^2\times S^1$ topology (Black Rings)
have been constructed \cite{Emparan:2001wn}.

In this letter we
will consider the  topology of  event horizons in dimensions
higher than four.
First, we reconsider Hawking's theorem \cite{HE}
and show it continues to hold in 
higher dimensions.
Using this and Thurston's geometric types
classification of three-manifolds \cite{t,thomas}, we find that the 
only possible
geometric types of event horizons in five dimensions are $S^3$
and $S^2 \times S^1$ (for a related discussion see \cite{Cai:2001su}).
In six dimensions we use the requirement that
the horizon is cobordant to a four-sphere, Friedman's
classification of topological four-manifolds and Donaldson's results
on smooth four-manifolds,  and show that 
simply connected event horizons are homeomorphic to $S^4$ or $S^2\times S^2$. 
We find allowed non-simply connected event 
horizons $S^3\times S^1$ and  $S^2\times \Sigma_g$ ($\Sigma_g$ is
a genus $g$ Riemann surface), and event 
horizons with finite non-abelian first homotopy group, whose universal
cover  is  $S^4$.
Finally, we will discuss the classification
in dimensions higher than six.

The letter is organized as follows.
In section 2 we will reconsider the uniqueness theorem of  \cite{HE}.
Following the steps of the proof in four dimensions
we will find that it holds also for asymptotically
flat stationary black holes in dimensions higher than four.
In section 3, we will consider the condition that the
integrated Ricci scalar curvature with respect to the metric induced 
on the event horizon is positive and the requirement that
the horizon is cobordant to a sphere, and study their implications
for the possible topologies of the event horizons in dimensions higher 
than four.

\section{Hawking's Theorem Revisited}

In this section we will reconsider Hawking's
theorem 
determining the two-sphere topology of 
the event horizon
of four-dimensional asymptotically flat stationary black holes \cite{HE}.
Following the steps of the proof in \cite{HE}, 
we will find that also for asymptotically flat stationary black holes
in dimensions higher than four the integrated Ricci scalar 
curvature  $\hat{\cR}$ with respect to the induced  metric $\hat{h}$
on the event horizon $\mh$, is positive
\begin{equation}
\int_{\mh}\hat{\cR} d\hat{S} > 0 \ .
\label{integral}
\eeq
The idea in the proof
is to use the fact that the shear and divergence are zero
at the horizon, but the divergence is positive outside the horizon.
In the next section we will study the implications of (\ref{integral}) 
for the topology of the event horizon for higher-dimensional asymptotically
flat stationary black holes.

Consider a stationary $n$-dimensional space-time $\cM$ with a metric $g$.
Stationary means that there exists a one-parameter group
of isometries whose orbits are time-like curves.
$\cM$ is required to be regular predictable, i.e. 
its future is predictable from a Cauchy surface \cite{HE}.

Denote by $(Y_1,Y_2)$ two future-directed null vectors orthogonal
to  $\mh$, normalized as 
\beq
Y_1^aY_{2a}=-1 \ .
\eeq
We take $Y_1$ to be the future-directed null vector pointing out of the horizon, 
and $Y_2$ to be the vector pointing into the horizon.
This still leaves us with the freedom to rescale 
\beq
Y_1\rightarrow e^{y}Y_1,Y_2 \rightarrow e^{-y}Y_2 \ .
\label{rescale}
\eeq
The (positive definite) induced metric on the horizon reads
\beq
\hat{h}_{ab}=g_{ab}+
Y_{1a}Y_{2b}+Y_{2a}Y_{1b} \ .
\eeq

We now deform the event horizon  by moving each point on it
a parameter distance $\omega$ along an orthogonal null
geodesic with tangent vector $Y_2^a$.
Following the same steps as in \cite{HE}
one derives the equation
\begin{equation}
\frac{d\hat{\theta}}{dw}
= p_{b;d} \hat{h}^{bd}
-\cR_{ac} Y_{1}^{a} Y_{2}^{c} +\cR_{adcb} Y_{1}^{d} Y_{2}^{c} Y_{2}^{a} Y_{1}^{b} 
+p_{a} p^{a} -Y_{1\,;c}^{a} \hat{h}_{\, d}^{c} Y_{2\,;b}^{d} \hat{h}_{\, a}^{b} \ ,
\label{eq:ConstraintV2}
\end{equation}
where
 $\hat{\theta} \equiv Y_{1\,;b}^{a} \hat{h}_{\, a}^{b}$
and 
$p^{a}= -\hat{h}^{ab} Y_{2c;b} Y_{1}^{c}$. 
 Note that the last term on the RHS is 
zero on the horizon, as the shear and the divergence of the null
geodesics with tangent vector $Y_1$ are zero there.

Using (\ref{rescale}), 
 $p^{a}\rightarrow p'^{a}=p^{a}+\hat{h}^{ab}y_{;b}$ and we get 
\begin{equation}
\left.\frac{d\hat{\theta}'}{dw'}\right|_{w=0} =p_{b;d} \hat{h}^{bd} +y_{;bd} \hat{h}^{bd}
-\cR_{ac} Y_{1}^{a} Y_{2}^{c} + \cR_{adcb} Y_{1}^{d} Y_{2}^{c} Y_{2}^{a} Y_{1}^{b} 
+p'_{a} p'^{a} \ .
\label{eq:ConstraintV3}
\end{equation}
Since $y_{;bd}\hat{h}^{bd}$ is the Laplacian of $y$ ($*d*d y$) on the 
$\left(n-2\right)$-dimensional horizon, we can use a 
theorem of Hodge to set the first four terms on the RHS equal to a constant by 
a particular choice of $y$, as follows:
The other three terms ($p_{b;d}\hat{h}^{bd} 
-\cR_{ac}Y_{1}^{a}Y_{2}^{c} +\cR_{adcb}Y_{1}^{d}Y_{2}^{c}Y_{2}^{a}Y_{1}^{b}$) are a 
0-form on the horizon, and their Hodge dual is a top-form. 
Any top-form $\phi$ on a connected manifold can be written as 
\beq
\phi = c \omega 
+ d\psi \ ,
\eeq
where $\omega$ is the volume form (normalized to unit integral) and $c$ 
is the integral of $\phi$ over the manifold. The theorem of Hodge states that for 
any form $\psi$, one can always find a form $u$ such that $d*d u = d\psi$. In this 
case, we set $y =-u$. Then the sum of the first four terms on the RHS is 
\begin{equation}
c = \int_{\mh} ( p_{b;d} \hat{h}^{bd} 
-\cR_{ac} Y_{1}^{a} Y_{2}^{c} +\cR_{adcb} Y_{1}^{d} Y_{2}^{c} Y_{2}^{a} Y_{1}^{b} ) 
d\hat{S} \ .
\label{eq:PositiveTerm1}
\end{equation}
The first term in the integral does not contribute because it is 
a divergence. 

The Gauss-Godazzi equations, evaluated on the horizon (where the shear 
and divergence are zero) yield 
\begin{equation}
\hat{\cR}= \cR_{ijkl} \hat{h}^{ik} \hat{h}^{jl}= \cR-
2\cR_{ijkl} Y_{1}^{i} Y_{2}^{j} Y_{1}^{k}
Y_{2}^{l} +4\cR_{ij} Y_{1}^{i} Y_{2}^{j} \ ,
\label{eq:Rhat}
\end{equation}
where $\hat{R}$ is the Ricci scalar associated to the induced metric, and the 
unhatted quantities are the curvature tensors of the full metric. The integral is 
therefore equal to
\begin{equation}
\int_{\mh} \left(-\frac{1}{2}\hat{\cR}+ \frac{1}{2}\cR+ \cR_{ab} 
Y_{1}^{a} Y_{2}^{b} \right) d\hat{S} \ .
\label{eq:PositiveTerm2}
\end{equation}
From the Einstein equations and the normalization $Y_1 \cdot Y_2 =-1$ we have 
\begin{equation}
\frac{1}{2}\cR +\cR_{ab} Y_{1}^{a} Y_{2}^{b} =8\pi T_{ab} Y_{1}^{a} Y_{2}^{b} \geq 0\ ,
\label{eq:Einstein}
\end{equation}
where we used the dominant energy condition. Thus we have:
\begin{equation}
\left.\frac{d\hat{\theta}'}{dw'}\right|_{w=0}=
\int_{\mh} (-\frac{1}{2}\hat{R}+ 8\pi T_{ab} Y_{1}^{a} Y_{2}^{b}) 
d\hat{S} +p'_{a} p'^{a} \ .
\label{eq:ConstraintV4}
\end{equation}

Suppose $\left.\frac{d\hat{\theta}'}{dw'}\right|_{w=0}$ is positive everywhere 
on the horizon. We then take $w'$ to be a small negative value, thereby looking
at a surface slightly outside the horizon on which $\hat{\theta}'$ is now negative. 
Such a surface is an outer trapped surface, which is forbidden in a stationary
regular predictable space-time satisfying the energy conditions. This is because the 
area of the light-cone of such a surface 
always shrinks in any time evolution, and hence cannot intersect future null infinity $\ScrIP$ 
(where the area would be infinite). Any region not observable from $\ScrIP$ is by definition within
the event horizon (and hence not "outer").

If 
\beq
\int_{\mh} (-\frac{1}{2}\hat{\cR} +8\pi T_{ab}Y_{1}^{a}Y_{2}^{b} )
d\hat{S} \ ,
\eeq
is positive, then it is possible to choose $y$ such that 
$\left.\frac{d\hat{\theta}'}{dw'}\right|_{w=0}$ is positive everywhere on the 
horizon, since $p'$ lies on the horizon, and hence is a space-like vector 
with positive (length)$^2$. This leads to an outer trapped surface. Thus, 
this quantity must be negative or zero. The dominant energy condition 
$T_{ab}Y_{1}^{a}Y_{2}^{b} \geq 0$
implies then that 
\begin{equation}
\int_{\mh}\hat{\cR} d\hat{S} \geq 0 \ .
\end{equation}

If this integral equals zero, then in order to avoid outer trapped 
surfaces, $T_{ab}Y_{1}^{a}Y_{2}^{b} = 0$.
Thus, the sum of the first four terms on the RHS of 
(\ref{eq:ConstraintV3}) equals zero, and 
\begin{equation}
p^{' a}{}_{;b} \hat{h}^b{}_a + \cR_{abcd} Y_1{}^a Y_2{}^b Y_1{}^c Y_2{}^d =0 \ ,
\label{eq:ZeroCurvCase}
\end{equation}
on $\mh$. Moreover, $p^{'a}$ must be zero on 
the horizon since $p^{'a} p'{}_a$ is positive-definite. This implies that each 
term in (\ref{eq:ZeroCurvCase}) must vanish independently on the horizon.
We can then choose the rescaling parameter $y$ such that $p^{'a}{}_{;b} - 
\frac{1}{2} \hat{R} =0$ on the deformed horizon for small negative 
$w'$. This gives rise to a marginally outer trapped surface, which is also forbidden.

Therefore, we arrive at the requirement (\ref{integral}),
for black hole horizons in asymptotically flat space of any dimensionality. 
Note, that for the theorem to hold we used the asymptotic flatness, and 
it does not hold for instance in asymptotically AdS spaces where
the dominant energy condition does not hold.

\section{The Topology of Event Horizons in Higher Dimensions}

In the previous section we found that for asymptotically
flat stationary black holes, in both the four-dimensional
case and higher dimensions, the integrated Ricci scalar 
curvature with respect to the metric induced
on the event horizon $\mh$ is positive (\ref{integral}).
When working in four dimensions, where the event horizon is a two-dimensional
manifold, this integral is proportional
to the Euler characteristic of the horizon manifold
and implies that the topology of
the event horizon is that of the two-sphere $S^2$.

In addition to (\ref{integral})
there is another constraint on the topology of the event horizon as
a consequence of the so-called topological censorship
\cite{Friedman:1993ty}.
Mathematically it requires that horizon is cobordant to a sphere
via a simply connected cobordism. For a $d$-dimensional horizon it means
that there is a simply connected $(d+1)$-dimensional oriented manifold
whose boundary is the oriented disjoint union $\mh \cup S^d$.
When working in four dimensions, topological censorship also implies that 
topology of the event horizon is the two-sphere $S^2$.

In the following we will use these two conditions and study their implications
for the topology of the event horizons in dimensions higher than four.

\subsection{Five-Dimensional Black Holes}

Consider five-dimensional stationary black holes.
Now the horizons are three-manifolds.
Thurston introduced eight geometric types in the classification of 
three-manifolds \cite{t} (see also \cite{thomas})
\footnote{This classification is called ``Thurston Geometrization Conjecture'',
and is claimed to have been established by Perelman.}. 
According to this classification
there are eight basic homogeneous geometries, up to an equivalence 
relation, called geometric types.
Out of these types one constructs geometric structures, which
are spaces that admit a complete
locally homogeneous metric 
\footnote{$M$ is called locally homogeneous if for any two points $x,y$
in $M$ there are neighborhoods
of these two points $U_x, U_y$ and an isometry that maps 
$(x,U_x)$ to $(y,U_y)$.}.
Any compact and oriented
three-manifold has a decomposition as a connected sum of
these basic geometric types.

We consider an orientable, connected, complete and  
simply connected Riemannian three-manifold $X$ 
which is homogeneous with respect to an
orientation preserving group of isometries $G$.
The eight geometric types classify $(X,G)$. 
The equivalence relation $(X,G) \sim (X',G')$ holds when
there is a diffeomorphism of $X$ onto $X'$, which takes the action of $G$ onto
the action of $G'$.
Out of these types one constructs spaces 
(geometric structures)
$M \simeq X/\Gamma$ where $\Gamma$ is a subgroup of $G$. 
Here the action of $\Gamma$
is discontinuous, discrete and free. 
$M$ is locally homogeneous with respect to the metric on $(X,G)$.
It is
isometric to the quotient of $X$ by $\Gamma$.
 
The first three types in the classification
are based on the three constant curvature spaces, the 3-sphere $S^3$ 
(Spherical
geometry),
which has a positive scalar curvature $\cR >0$
and isometry group $G=SO(4)$, the Euclidean space  $R^3$  (Euclidean geometry)
with  $\cR = 0$ and 
isometry group $G=R^3 \times SO(3)$ and
the hyperbolic space $H^3$  (Hyperbolic geometry)
with  $\cR < 0$ and 
isometry group $G=PSL(2,C)$.
Of these three geometric types, only the $S^3$ type 
satisfies the condition (\ref{integral})
and is allowed as an horizon.

The next two types are based on $S^2 \times R$ and  $H^2 \times R$.
Of these two  geometric types, only  the
$S^2 \times R$  type satisfies the condition (\ref{integral}) and
is allowed as an horizon.
In this allowed class we have $S^2\times S^1$.

The last three geometric types are Nil geometry, Sol geometry
and the universal cover
of the Lie group $SL(2,R)$.

\vspace*{5mm}

{\it The Nil geometry}: this is the geometry of the 
three-dimensional Lie group of $3\times 3$ real upper triangular
matrices of the form
\begin{eqnarray}\label{matrix}
\left(\begin{array}{ccc} 1 & x & z\\
                         0 & 1 & y\\
                         0 & 0 & 1 \end{array}\right) 
\end{eqnarray}
under matrix multiplication (Heisenberg group).
We can think about Nil as $(x,y,z)\in R^3$ with
the multiplication
\beq
(x,y,z)\cdot (x',y',z') = (x+x', y+y',z+z'+x y') \ .
\eeq

The Nil  metric (the left-invariant metric on $R^3$) is given by
\beq
ds^2 = dx^2 + dy^2  +  (dz -x dy)^2 \ ,
\eeq
and has $\cR = -\frac{1}{2}$.
The Nil geometry type does
not satisfy the condition (\ref{integral}) and
is not allowed as an horizon.

\vspace*{5mm}
{\it The Sol geometry}: this is the geometry of the 
of the Lie group  obtained
by the semidirect product of $R$ with $R^2$.
We can think about Sol as $(x,y,z)\in R^3$ with
the multiplication
\beq
(x,y,z)\cdot (x',y',z') = (x+ e^{-z}x', y+ e^{z}y',z+z') \ .
\eeq
The left-invariant Sol metric is given by
\beq
ds^2 = e^{2 z}dx^2 + e^{-2 z}dy^2+  dz^2 \ ,
\eeq
and has $\cR =-2$.
The Sol geometry type does
not satisfy the condition (\ref{integral}) and
is not allowed as an horizon.

\vspace*{5mm}
{\it The $\widetilde{Sl(2,R)}$ geometry}: this is the geometry of the 
universal covering
of the three-dimensional Lie group
of all $2\times 2$ real matrices with determinant
one $Sl(2,R)$.
 The $\widetilde{Sl(2,R)}$ geometry type has
$\cR < 0$. Thus, it 
 does
not satisfy the condition (\ref{integral}) and
is not allowed as an horizon.

We should note, that for the Nil, Sol and  $\widetilde{Sl(2,R)}$
geometries, we considered the natural metrics and have not attempted
to prove that there are no other differential structures and metrics
with different properties.

\vspace*{5mm}
{\it Summary}: 
We find that only two geometric types are
allowed  horizons in five dimensions: the $S^3$ geometric type
and the  $S^2\times R$ geometric type.
Indeed, black hole solutions of both geometric
types with compact event horizon topologies have been constructed, 
namely $S^3$ and 
$S^2\times S^1$.

\subsection{Six-Dimensional Black Holes}

Consider now six-dimensional stationary
black holes in asymptotically flat space-times. 
We will use 
topological censorship\footnote{In this section we will use
topological censorship as oriented cobordism of $\mh$ and $S^4$ without 
requiring the five-dimensional manifold to be simply connected.} together with Friedman's  classification of
four-manifolds and Donaldson's results
on smooth four-manifolds, in order to classify possible
event horizons.

First we note that oriented cobordism from the event horizon to a four-sphere $S^4$
exists if and only if the 
horizon manifold $\mh$ has vanishing Pontrjagin and Steifel-Whitney numbers \cite{mils}.
In general, two smooth closed $n$-dimensional manifolds are cobordant 
iff all their corresponding Steifel-Whitney numbers are equal.
If in addition we require the cobordism to be oriented
then (when $n=4k$) their corresponding Pontrjagin numbers are equal.
In the following we will study the restriction that these set 
on the topology of the four-manifold event horizons.

We start by considering simply connected event horizons, that is
\beq
\Pi_1(\mh) = 0 \ .
\eeq
This, in particular, implies that the cohomology groups
$H^1(\mh)$ (which is the abelianization of the first homotopy
group), and $H^3(\mh)$ (by Hodge duality) vanish.
One can use the second cohomology 
group $H^2(\mh)$ to define an intersection form
\beq
Q(\alpha,\beta) = \int_\mh \alpha\wedge \beta \ ,
\eeq
where $\alpha,\beta \in H^2(\mh)$.
$Q$ is the basic topological invariant of a compact
four-manifold.
Note that, since the four-sphere $S^4$ has zero second cohomology
group, all its intersection numbers vanish and
\beq
Q(S^4) = {\bf (0)} \ .
\eeq 

$Q$ is symmetric, non-degenerate with $rank(Q) = b_2 \equiv 
dim~H^2(\mh)$, and can be diagonalized over $R$.
The signature $\sigma$ of a four-manifold is defined
by the difference of positive and negative
eigenvalues
of $Q$. It can be expressed using the Hirzebruch signature
theorem as
\beq
\sigma(\mh) = \frac{1}{3}\int_{\mh}p_1 \ ,
\eeq
where $p_1$ is the first Pontrjagin class, which can be expressed
using the Riemann curvature
as
\beq
p_1 = -\frac{1}{8 \pi^2}Tr R\wedge R \ .
\eeq
Since $p_1(S^4)$ vanishes,
topological censorship implies then that  the signature of $Q(\mh)$
vanishes
\beq
\sigma(Q(\mh)) = 0 \ .
\label{sig}
\eeq
Mathematically, the signature is cobordant invariant.

Consider next the Stiefel-Whitney classes
\beq
\omega_i \in H^i(\mh, Z_2) \ .
\eeq
For a compact, simply-connected orientable manifold
$\omega_1=\omega_3=0$.
$\omega_2$ is the obstruction to a spin-structure.
Although $\omega_2(S^4)=0$, oriented cobordism does not imply
that 
the second  Stiefel-Whitney class  of  $\mh$ is zero.
In other words, 
$\mh$ is not necessarily a spin manifold.

The intersection form $Q$ is actually defined on the lattice $H^2(\mh, Z)$ and is
a unimodular ($det(Q) = \pm 1$) symmetric bilinear form over the integers.
One says that $Q$ is of even type if
\beq
Q(\alpha,\alpha) \in 2 Z
\eeq
for all $\alpha \in  H^2(\mh, Z)$.
If   $\omega_2=0$ then $Q$ is even (as implied by Wu's formula
\cite{mils}).
Thus, event horizons which are spin manifolds are characterized as
topological
four-manifolds 
by an intersection form $Q(\mh)$, which has vanishing signature and is
of even type.
When the event horizons are not spin manifolds,  $\omega_2\neq 0$ and
$Q(\mh)$ is odd.
In this case there are two topological four-manifolds $M_H$ for
a given intersection form. They are distinguished by the
Kirby-Siebenmann invariant, which is zero if $M_H\times S^1$ is
smooth and one if  $M_H\times S^1$ is not smooth.

In the following we will use combined results of Friedman's classification
and Donaldson's theorems (see \cite{fu,dk}).
In the classification of possible intersection forms of $\mh$
we
distinguish two cases:\\
(i) $Q(\mh)$ is positive definite,\\
(ii) $Q(\mh)$ is indefinite.\\
Consider first the case when  $Q(\mh)$ is positive definite.
If  $Q(\mh)$ is even then $\mh$ is homeomorphic to the
four-sphere $S^4$ (see \cite{fu} Corollary (2.27)).
If $Q(\mh)$ is odd then  $\mh$ is homeomorphic to a connected
sum of $CP^2$'s. However, since $Q(CP^2) = (\bf{1})$, the signature of
the connected sum in nonzero, and this is not an allowed horizon.

If  $Q(\mh)$ is indefinite, then if it is even it can be written
as (Hasse and Minkowski)
\beq
Q(\mh) = a E_8 + b H,~~~~a,b \in Z~~~b\neq 0 \ ,
\eeq
where
 $E_8$ is the Cartan matrix of
the Lie algebra $E_8$ and
\begin{eqnarray}\label{matrix}
H = \left(\begin{array}{cc} 0 & 1 \\
                        1 & 0 \end{array}\right) \ ,
\end{eqnarray}
is the intersection
form of $S^2\times S^2$.
Since the signature $\sigma(E_8) = 8$ and we require
that $\sigma(Q(\mh))= 0$
this implies that $a=0$. The basic case is $b=1$ and $\mh$ is
$S^2\times S^2$.
If we take $b>1$ we will get 
a connected sum
of $S^2\times S^2$.

We should note, however, that by using connected sums there is a way
to construct other event horizons whose
 intersection form has vanishing signature and is
of even type.
Consider for instance a $K3$. Its
intersection form is $Q(K3) = -2 E_8 + 3 H$. Since its
signature
is nonzero, $K3$ is not an allowed horizon. However, if we take a
connected
sum  of the $K3$ and $-K3$, where $-K3$ has an opposite orientation
we get an even intersection form with vanishing signature, since
 $Q(-K3)= -Q(K3)$. 

If  $Q(\mh)$ is indefinite and odd then $\mh$ is a connected sum of
$\pm CP^2$'s, where $-CP^2$ has the  opposite orientation
of $CP^2$ and  $Q(-CP^2)= -Q(CP^2) = (\bf{-1})$.

In the following we consider the possible event horizons up
to the connected sum operation.

\vspace*{5mm}
{\it Summary}: 
We found that if the horizon is simply connected
then it is homeomorphic to $S^4$ or to 
$S^2\times S^2$.
Note that both  $S^4$ and 
$S^2\times S^2$ are spin manifolds.
 
Consider next the case when $\mh$ is not simply connected.
When $Q(\mh)$ is positive definite, one can relax the condition that
 $\Pi_1(\mh)=0$ by requiring only that there are no non-trivial 
homomorphisms of $\Pi_1(\mh)$ into $SU(2)$. This implies that every flat 
$SU(2)$ bundle over $\mh$ is trivial and that
$H_1(\mh)$ being the abelianization of $\Pi_1(\mh)$ vanishes \cite{fu}.
This allows $\Pi_1(\mh)$ to be any finite simple nonabelian group.
With this relaxed condition we get four-manifold event horizons,
whose universal cover is $S^4$.

There are three other non-simply connected cases that we would like to explore.
First, consider  $T^4$. Its intersection form is 
$Q(T^4) = 3H$ and it is not excluded by the previous discussion
from being 
an event horizon.
However, it does not satisfy our curvature condition (\ref{integral}).
Next consider $S^3\times S^1$. It is not ruled out by our analysis
since it has vanishing  Pontrjagin and Steifel-Whitney numbers.
Also, it satisfies (\ref{integral}).
The last examples are $\Sigma_g\times \Sigma_h$, where
$\Sigma_g$ and $\Sigma_h$ are Riemann surfaces of genus $g$ and $h$
respectively, and we have assumed that the induced metric decomposes
as a direct (unwarped) sum.
Assuming a product metric, the condition (\ref{integral}) reads
\beq
(g-1)Vol(\Sigma_h) + (h-1)Vol(\Sigma_g) < 0 \ .
\eeq
This can be satisfied by $\Sigma_h = S^2$ and 
\beq
g < 1 + \frac{Vol(\Sigma_g)}{Vol(S^2)} \ .
\label{volume}
\eeq
We cannot exclude, a priori, the possibility that the ratio
of volumes in (\ref{volume}) can be as large as we want
and therefore all genera $g$ are allowed.
We  encountered above the case $g=0$, namely the horizon
$S^2\times S^2$.
The intersection
form $Q(S^2\times \Sigma_g)= H$ and it is not ruled out by topological censorship.

\vspace*{5mm}
{\it Summary}: 
We found that if the event horizon has vanishing  first homotopy group
then it is homeomorphic to $S^4$ or $S^2\times S^2$. If the event horizon has 
finite simple nonabelian first homotopy group and positive intersection form, 
then its universal cover is homeomorphic to $S^4$.
We found other allowed non-simply connected cases  $S^3\times S^1$ with
first homotopy group $Z$ and $S^2\times \Sigma_g$ with
first homotopy group\footnote{
$\Pi_1(\Sigma_g)$ is generated by $a_1,b_1,...,a_g,b_g$ with
the relation $a_1b_1a_1^{-1}b_1^{-1}...a_gb_ga_g^{-1}b_g^{-1}=1$.}
$\Pi_1(\Sigma_g)$. 

\subsection{Comments on Higher Dimensions}

In the following we will
make some comments on the classification of the event
horizons above six dimensions. 
The event horizons $\mh$ are now closed differentiable $n$-manifolds
with dimension $n$ higher than four, cobordant to the n-sphere $S^n$.
If $\mh$ is homotopic to  $S^n$, then by the generalized Poincare 
conjecture (proven when $n>4$) $\mh$ is homeomorphic to $S^n$ \cite{smale}.

An important concept in differential topology
is that of h-cobordism.
Two cobordant $n$-manifolds are h-cobordant if their inclusion map
in the $n+1$-dimensional manifold are homotopy equivalent.
The  h-cobordism theorem \cite{smale}
implies that if the 
horizon manifold $\mh$ is h-cobordant to $S^n$ then it is diffeomorphic 
to $S^n$. Note, however, that h-cobordism is a stronger requirement
than what is implied by topological censorship.

Let us introduce the concept of spin cobordism.
Mathematically, it
requires in our context the existence of an $(n+1)$-dimensional compact
spin manifold, whose boundary is the oriented disjoint union of $S^n$
and $M_H$. 
In particular,  $M_H$ is a spin manifold whose spin structure
is induced from that of the  $(n+1)$-dimensional manifold.
The concept of spin cobordism may be
relevant, since we are mainly interested in
higher-dimensional
black holes solutions to supergravity equations as the low energy effective
description of the superstring equations. Therefore, we would like
the geometry to accommodate fermions.
Note, that in the classification of the previous section,
  spin cobordism  would have implied
that
the intersection form is of even type.

There are several useful results that help in the classification of
the event horizons of dimensions higher than four:
\begin{itemize}
\item{}{\it $n=5$} :
For five-dimensional manifolds with vanishing second Steifel-Whitney 
class , there exists a classification of all possible
closed simply connected manifolds \cite{smale}.
The manifolds are in 1-1 correspondence with
finitely generated abelian groups.

\item{}
{\it $n=6$} :
Six-dimensional closed manifolds with vanishing first and second
homotopy groups $\Pi_1=0$ and $\Pi_2=0$ (2-connected) are homeomorphic
to $S^6$ or connected sum of copies of $S^3\times S^3$  \cite{smale}.

\item{}
{\it $n=2k$} : There are general results which enumerate
the $(k-1)$-connected $2k$-manifolds (Wall) \cite{smale}.

\item{} {\it $n \geq 5$} : If $M_H$ is $n$-dimensional compact, simply
connected and spin cobordant to
$S^n$, it is obtained from $S^n$ by doing surgery on spheres of
codimension greater than two \cite{GL}.

\end{itemize}
                    
We will leave the complete analysis of the possible topologies of
event horizons of stationary black holes in asymptotically flat 
space-times with $dim~ \mh > 4$ to the future.

\section*{Acknowledgements}

We would like to thank J. Distler, D. Freed and B. Kol
for valuable discussions.

\end{document}